\documentclass{amsart}
\usepackage{amsmath}
\usepackage[misc]{ifsym}
\usepackage[colorlinks=true]{hyperref}
\hypersetup{urlcolor=blue, citecolor=red}
\allowdisplaybreaks

\usepackage{graphicx}
\usepackage{tikz}
\usetikzlibrary{decorations.markings}
\usepackage{subfigure}
\usepackage{caption}

\newtheorem{theorem}{Theorem}[section]

\newtheorem{proposition}[theorem]{Proposition}

\theoremstyle{definition}
\newtheorem{definition}[theorem]{Definition}

\title[Bistability and chaos in the discrete two-gene model]{Bistability and chaos in the discrete two-gene Andrecut-Kauffman model} 

\author[M. Rosman, M. Palczewski, P. Pilarczyk and A. Bartłomiejczyk$^*$]{}


\thanks{$^\dag$ Both authors contributed equally to the research}
\thanks{$^*$Corresponding author, \url{https://pg.edu.pl/p/agnieszka-bartlomiejczyk-13028} \\[12pt]
This is the author accepted manuscript (postprint) of the paper published
in \textit{Discrete and Continuous Dynamical Systems -- Series B} (DCDS-B),
Vol.\ 30, No.\ 11, November 2025, pp.\ 4442--4461,
\url{https://doi.org/10.3934/dcdsb.2025028}. \\
This postprint is licensed under the \href{https://creativecommons.org/licenses/by-nc-nd/4.0/}%
{CC-BY-NC-ND 4.0 Creative Commons License}.
}

\begin{document}
\maketitle

\centerline{\scshape
Mikołaj Rosman$^{\dag 1}$
Michał Palczewski$^{\dag 1,2}$
Paweł Pilarczyk$^{1,3}$
and Agnieszka Bartłomiejczyk$^{*1,4}$}

\medskip

{\footnotesize
 \centerline{$^1$Faculty of Applied Physics and  Mathematics,
Gda\'{n}sk University of Technology,
ul.\ Narutowicza 11/12,
80-233 Gda\'{n}sk, Poland}
} 

\medskip

{\footnotesize
 \centerline{$^2$Doctoral School,
Gda\'{n}sk University of Technology,
ul.\ Narutowicza 11/12,
80-233 Gda\'{n}sk, Poland}
}

\medskip

{\footnotesize
 \centerline{$^3$Digital Technologies Center,
Gda\'{n}sk University of Technology,
ul.\ Narutowicza 11/12,
80-233 Gda\'{n}sk, Poland}
}

\medskip

{\footnotesize
 \centerline{$^4$BioTechMed Center,
Gda\'{n}sk University of Technology,
ul.\ Narutowicza 11/12,
80-233 Gda\'{n}sk, Poland}
}

\bigskip


\begin{abstract}
We conduct numerical analysis of the 2-dimensional discrete-time gene expression model originally introduced by Andrecut and Kauffman (Phys. Lett. A 367: 281--287, 2007).
In contrast to the previous studies, we analyze the dynamics with different reaction rates $\alpha_1$ and $\alpha_2$ for each of the two genes under consideration.
We explore bifurcation diagrams for the model with $\alpha_1$ varying in a wide range and $\alpha_2$ fixed.
We detect chaotic dynamics by means of the positive maximum Lyapunov exponent and we scan through selected parameters to detect those combinations for which chaotic dynamics can be found in the model.
Moreover, we find bistability in the model, that is, the existence of two disjoint attractors.
Both situations are interesting from the point of view of applications, as they imply unpredictability of the dynamics encountered.
Finally, we show some specific values of parameters of the model in which the two attractors are of different kind (a periodic orbit and a chaotic attractor) or of the same kind (two periodic orbits or two chaotic attractors).
\end{abstract}

\section{Introduction}\label{sec:intro}

Modeling gene expression plays a key role in understanding the regulatory mechanisms within cells and the dynamics of biological processes at the molecular level. The development of mathematical models that accurately reflect the course of these processes allows for predicting the behavior of biological systems and contributes to the advancement of therapeutic strategies and genetic engineering. Gene expression models enable the study of complex networks of interactions between genes, transcripts, and proteins, which in turn contributes to a better understanding of phenomena such as cellular responses to stress and cell differentiation processes.

Despite the undeniable complexity of the chemical reactions and kinetics involved in transcription and translation processes, the expression of a single gene does not reflect the intricacies of processes occurring within a living cell. When more genes are present in a system, their corresponding proteins can act as transcription factors, binding to different promoters, creating an extensive network of dependencies. Due to their significance for all biological processes, numerous attempts have been made over the past decades to accurately model gene regulatory networks.

The first approach to modeling gene expression involves continuous-time models, in which the dynamics is described using systems of differential equations. This type of modeling allows for capturing the variables that describe the concentrations of gene products, such as mRNA or proteins, in a smooth manner over time. Among the classical models of this type are systems of differential equations. These systems form the basis for analyzing dynamic dependencies among various components of a biological system, such as Goodwin's model (see \cite{goodwin-1965}) and the Michaelis-Menten model (see \cite{michaelis-menten-1913}), which have been further developed by many authors. Such models constitute effective tools for studying kinetic reactions and gene transcription, although their limitation is the necessity of assuming an immediate response of the system to changes in parameters. Therefore, in order to better capture biological realities, models that incorporate time delays are considered.
Delay models take into account time shifts that reflect the actual phenomena taking place during the synthesis of proteins and mRNA. This approach allows for a more precise reflection of processes where natural delays exist between the initiation of transcription and protein synthesis.

The introduction of time delays into gene expression models has been proposed in numerous studies. For example, the authors of \cite{bartlomiejczyk-bodnar-2020,bartlomiejczyk-bodnar-2023,bodnar-bartlomiejczyk-2012} analyze continuous mathematical models of Hes1 gene expression, incorporating various aspects such as time delays, feedback mechanisms, and dimerization processes. The paper \cite{bodnar-bartlomiejczyk-2012} introduces separate delays for transcription and translation, highlighting the impact of the total delay on stability and the occurrence of a Hopf bifurcation. In \cite{bartlomiejczyk-bodnar-2020}, the model is expanded to include different time scales and is reduced to the classical Hes1 model, examining equilibrium stability and dynamics. Finally, further development of the model is proposed in \cite{bartlomiejczyk-bodnar-2023} by including dimerization, and analysis of the impact of delays on bifurcations and stability of limit cycles is being conducted, which results in deeper insight into oscillations in Hes1 gene expression. Similarly, the authors of \cite{piotorwska-bartlomiejczyk-bodnar-2018} investigate a generalized p53-Mdm2 model and demonstrate that oscillations in gene expression can arise with or without time delays, depending on specific stability conditions and bifurcation criteria.

The groundwork for the model analyzed in our research was laid out by Kauffman in 1969 in the form of Boolean networks \cite{KAUFFMAN1969437}. A Boolean network is a finite set of Boolean variables changing in discrete time steps according to predetermined rules. A single network can have several distinct modes of behavior, by analogy to one genome creating specialized cell types. Further analysis of random Boolean networks was facilitated in 2005 when Andrecut used mean field theory to calculate probabilities of node states in a network given only global parameters. Moreover, he has calculated Lyapunov exponents of such networks as a function of connectivity, and has shown that for the purpose of investigating chaotic behavior the discrete network can be replaced with a continuous map. This allowed for application of Sharkovskii’s theorem to find possible orbits present in the model \cite{Andrecut_2005}.

In 2006, Andrecut and Kauffman proposed a continuous model in which individual reactions between genes have been replaced with their average results, reducing computational complexity and making the outcome deterministic, as opposed to the more realistic stochastic simulations. The regulatory network is represented as a system of differential equations, with exact formulae derived directly from the underlying chemical reactions. The model accounts for \(N\) genes and protein multimers up to length \(n\), including mixed protein chains. Fourier analysis of numerical solutions has shown that this model admits a range of different behaviors, such as stationary solutions, periodic orbits, and chaotic behavior as well \cite{Andrecut_2006}.

A more approachable, discrete-time model was proposed a year later in \cite{andrecut_main}. The number of genes was fixed at $2$ and only homo-multimers of length exactly \(n\) were taken into consideration. This led to a system of two mutually dependent equations. Despite the considerable simplification, the new model seems to represent the complex dynamics of the original model well, displaying different modes of behavior under various conditions. Combined with the relative ease of simulating a discrete model, this has encouraged further research.

In our work, we focus on the discrete Andrecut-Kauffman model. However, we consider a generalized form in which a different reaction rate may be associated to each gene (see Section~\ref{sec:model}). This distinction allows for a more precise representation of gene expression dynamics. Specifically, instead of a single, general rate $\alpha$, inclusion of $\alpha_1$ and $\alpha_2$ in the equations of the model enables capturing various regulatory mechanisms that may influence transcription processes. Additionally, we incorporate the Lyapunov exponent into our analysis, which enables us to measure the model's stability and sensitivity to initial conditions, an essential factor for understanding the complex dynamics of discrete models in a biological context.

The purpose of our research is to investigate the dynamics encountered in the generalized discrete Andrecut-Kauffman model important for applications. We put emphasis on the existence of chaotic dynamics, and also on the phenomenon of bistability, that is, the existence of two distinct attractors, possibly of different types, such as an attracting periodic orbit and a chaotic attractor. Both phenomena are of undeniable importance from the biological point of view. Chaotic dynamics implies unpredictability of the long-term behavior of the modeled system. Bistability implies the fact that the system can stabilize in two different ways depending on the initial conditions.

The paper is organized as follows. In Section~\ref{sec:mechanisms}, we explain the gene expression mechanisms that yield the difference equations that we analyze. The equations are introduced in Section~\ref{sec:model}. Then in Section~\ref{sec:absorbing}, we establish an absorbing set that captures all the bounded dynamics. In Section~\ref{sec:chaoticperiodic}, we show evidence that admitting different reaction rates proposed in Section~\ref{sec:equations} can result in qualitative change of dynamics and is thus an important direction of investigation. In Section~\ref{sec:bifdiag}, we show and analyze bifurcation diagrams of the model with one of the reaction rates varying in a wide range while keeping the other rate fixed. Then we introduce a numerical procedure for the computation of the maximum Lyapunov exponent in our $2$-dimensional model in Section~\ref{sec:LyapComp}, and we apply it to the search for chaotic dynamics in Section~\ref{sec:posLyap}. We then analyze the maximum Lyapunov exponents for a wide range of reaction rates in Section~\ref{sec:plotLyap}. Then in Section~\ref{sec:asymmetry}, we discuss the asymmetry found in the computation of the maximum Lyapunov exponent, which yields to the detection of three kinds of bistability in the model that we discuss in Section~\ref{sec:bistability}.
The paper concludes with a discussion in Section~\ref{sec:conclusion}, where we address the importance of our findings for understanding gene expression dynamics, emphasizing the role of parameters in the stability of the analyzed model and its potential to exhibit chaotic behavior. We also suggest future research directions, such as the analysis of bifurcations, unstable fixed points, and global dynamics in the context of Systems biology and controlled genetic systems.

\section{Gene expression mechanisms}\label{sec:mechanisms}

Gene expression encompasses all the processes that determine the amount of a specific mRNA and protein produced within a cell. This series of processes includes two fundamental stages: transcription, where a gene's DNA sequence is transcribed into mRNA, and translation, where the mRNA is subsequently translated into a protein. Gene expression often serves as a foundational aspect in cell biology research, providing insight into mechanisms both at the microscopic and molecular scale. Understanding how gene expression is regulated and coordinated is crucial for grasping the complexity of cellular systems~\cite{larson}.

\subsection{Transcription}
\label{sec:transcription}
Transcription is the process in which enzymes use one strand of DNA within a gene as a template to produce messenger RNA (mRNA). RNA polymerase, aided by proteins called transcription factors, binds to a specific sequence within the gene known as the promoter and separates the two DNA strands. The template strand, or the antisense strand, is used to generate the mRNA, while the other strand is the nontemplate or sense strand. RNA polymerase initiates mRNA synthesis at the start codon without needing a primer and moves downstream along the gene in a process called elongation. It synthesizes the mRNA by reading the antisense strand and generating the mRNA, adding RNA nucleotides as it goes. This process is similar to how DNA polymerase synthesizes DNA, except that RNA is being synthesized with ribose instead of deoxyribose and uracil instead of thymine. Unlike DNA replication, RNA polymerase re-zips the DNA behind it, keeping only 10 to 20 bases exposed at a time. When RNA polymerase reaches the end of the gene, termination occurs, the enzyme detaches, and the DNA returns to its original state. The resulting mRNA, after a few modifications during RNA processing, leaves the nucleus and moves to the cytoplasm, where it finds a ribosome~\cite{alberts2002molecular}.

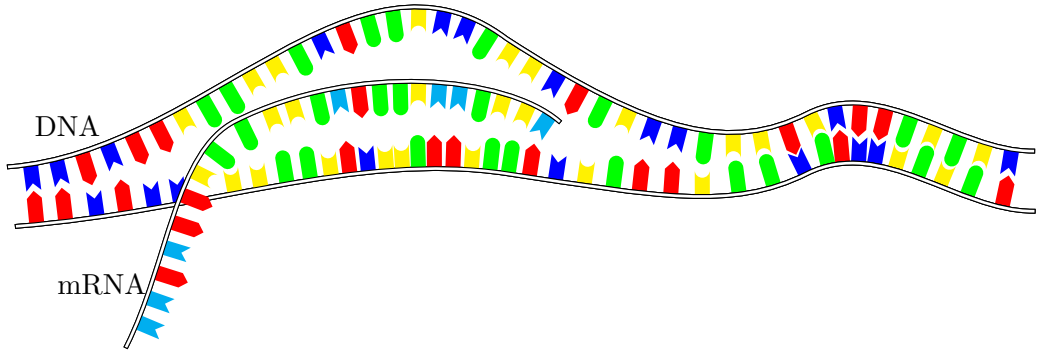
\begin{figure}[htbp!]
    \centering
    \resizebox{\textwidth}{!}{
    \begin{tikzpicture}[
        adenin/.style = {decoration = {markings,
        mark = at position #1 with { \adenin{(0, 0)}{270} }
  }}, thymin/.style = {decoration = {markings,
        mark = at position #1 with { \thymin{(0, 0)}{270} }
  }}, guanin/.style = {decoration = {markings,
        mark = at position #1 with { \guanin{(0, 0)}{270} }
  }}, cytosin/.style = {decoration = {markings,
        mark = at position #1 with { \cytosin{(0, 0)}{270} }
  }}, uracyl/.style = {decoration = {markings,
        mark = at position #1 with { \uracyl{(0, 0)}{270} }
    }}]
  \newcommand*{\adenin}[2]{\begin{scope}[shift = {#1}, rotate = #2, fill = red]%
        \fill(0, -.1) -- (.35, -.1) -- (.45, 0) -- (.35, .1) -- (0, .1) -- cycle;%
  \end{scope}}%
  \newcommand*{\thymin}[2]{\begin{scope}[shift = {#1}, rotate = #2, fill = blue]%
        \fill(0, -.1) -- (.35, -.1) -- (.25, 0) -- (.35, .1) -- (0, .1) -- cycle;
  \end{scope}}%
  \newcommand*{\guanin}[2]{\begin{scope}[shift = {#1}, rotate = #2, fill = green]%
        \fill(0, -.1) -- (.35, -.1) arc(-90:90:.1) -- (.35, .1) -- (0, .1) -- cycle;
  \end{scope}}%
  \newcommand*{\cytosin}[2]{\begin{scope}[shift = {#1}, rotate = #2, fill = yellow]%
        \fill(0, -.1) -- (.35, -.1) arc(270:90:.1) -- (.35, .1) -- (0, .1) -- cycle;%
  \end{scope}}%
  \newcommand*{\uracyl}[2]{\begin{scope}[shift = {#1}, rotate = #2, fill = cyan]%
        \fill(0, -.1) -- (.35, -.1) -- (.25, 0) -- (.35, .1) -- (0, .1) -- cycle;
  \end{scope}}%
     \draw[double distance = 1pt, line cap = rect,
        thymin/.list = {5pt, 15pt, 35pt,125pt,175pt,185pt},
        adenin/.list = {25pt, 45pt, 55pt,135pt},
        guanin/.list = {75pt, 85pt, 115pt,145pt,155pt,195pt},
        cytosin/.list = {65pt,95pt,105pt,165pt,205pt,215pt},
      preaction = {decorate}, shorten <= -3pt]
        (-.5, -.2) .. controls (2, 0) and (4, 3) .. (6, 1.5)
        .. controls (8, 0.2)  and (9, 0) .. (10, 0.5)
        .. controls (11, 1) and (12, 0) .. (13, 0);

    \draw[double distance = 1pt, line cap = rect,
        thymin/.list = {225pt,265pt,275pt},
        adenin/.list = {235pt,315pt},
        guanin/.list = {245pt,285pt},
        cytosin/.list = {255pt,295pt,305pt,325pt},
      preaction = {decorate}, shorten <= -3pt]
        (-.5, -.2) .. controls (2, 0) and (4, 3) .. (6, 1.5)
        .. controls (8, 0.2)  and (9, 0) .. (10, 0.5)
        .. controls (11, 1) and (12, 0) .. (13, 0);

    \draw[double distance = 1pt, line cap = rect,
        thymin/.list = {335pt,405pt},
        adenin/.list = {345pt,355pt},
        guanin/.list = {365pt,385pt},
        cytosin/.list = {375pt,395pt},
      preaction = {decorate}, shorten <= -3pt]
        (-.5, -.2) .. controls (2, 0) and (4, 3) .. (6, 1.5)
        .. controls (8, 0.2)  and (9, 0) .. (10, 0.5)
        .. controls (11, 1) and (12, 0) .. (13, 0);
        
     \draw[double distance = 1pt, line cap = rect,
        thymin/.list = {60pt,66pt},
        adenin/.list = {9pt,72pt},
        guanin/.list = {23pt,43pt,77.5pt},
        cytosin/.list = {34pt,52pt},
      preaction = {decorate}, shorten <= -3pt]
        (12.9, -0.8) .. controls (12, -0.8) and (11, 0.2) .. (10, -0.3)
        .. controls (9, -0.8)  and (8, -0.6) .. (6, -0.3)
        .. controls (4, 0) and (2, -0.8) .. (-0.5, -1);

     \draw[double distance = 1pt, line cap = rect,
        thymin/.list = {86.5pt,183pt},
        adenin/.list = {139pt,151pt,193pt},
        guanin/.list = {99.5pt,113pt,161pt},
        cytosin/.list = {127pt,172pt},
      preaction = {decorate}, shorten <= -3pt]
        (12.9, -0.8) .. controls (12, -0.8) and (11, 0.2) .. (10, -0.3)
        .. controls (9, -0.8)  and (8, -0.6) .. (6, -0.3)
        .. controls (4, 0) and (2, -0.8) .. (-0.5, -1);

     \draw[double distance = 1pt, line cap = rect,
        thymin/.list = {254pt,326pt,336pt,358pt},
        adenin/.list = {222pt,229pt,261pt,347pt,370pt,381pt},
        guanin/.list = {201pt,208pt,235pt,276pt,285pt},
        cytosin/.list = {215pt,241pt,247pt,268pt,295pt,305pt},
      preaction = {decorate}, shorten <= -3pt]
        (12.9, -0.8) .. controls (12, -0.8) and (11, 0.2) .. (10, -0.3)
        .. controls (9, -0.8)  and (8, -0.6) .. (6, -0.3)
        .. controls (4, 0) and (2, -0.8) .. (-0.5, -1);

    \draw[double distance = 1pt, line cap = rect,
        uracyl/.list = {134pt,173pt,181pt,217pt},
        adenin/.list = {142pt},
        guanin/.list = {95pt,126pt,150pt,158pt,190pt},
        cytosin/.list = {107pt,116pt,165pt,198pt,208pt},
      preaction = {decorate}, shorten <= -3pt]
        (1, -2.5) .. controls (1.5, -1.5)  and (1.6, 0) .. (2.4, 0.4)
        .. controls (4, 1.2) and (6, 1) .. (6.7, 0.4);

    \draw[double distance = 1pt, line cap = rect,
        uracyl/.list = {7pt,17pt,37pt},
        adenin/.list = {27pt,47pt,58pt},
        guanin/.list = {79pt},
        cytosin/.list = {68pt},
      preaction = {decorate}, shorten <= -3pt]
        (1, -2.5) .. controls (1.5, -1.5)  and (1.6, 0) .. (2.4, 0.4)
        .. controls (4, 1.2) and (6, 1) .. (6.7, 0.4);
        
\draw (-0.3,0.5) node [anchor=north west][inner sep=0.75pt]   [align=left] {DNA};

\draw (0,-1.6) node [anchor=north west][inner sep=0.75pt]   [align=left] {mRNA};

\end{tikzpicture}
}
     \caption{Schematic representation of DNA transcription.}   
    \label{fig:trascription}
\end{figure}

\subsection{Translation}
\label{sec:translation}
Translation occurs in the ribosome, where mRNA serves as a code for a specific protein. Each set of three bases on the mRNA, called codons, encodes a specific anticodon carried by a transfer RNA (tRNA), each linked to a particular amino acid. The arrangement of nucleotides into codons is called the reading frame. With four bases to choose, and codons consisting of three bases each, there are $64$ possible codons, enough to encode all the necessary amino acids. Each codon corresponds to a particular amino acid; AUG is the start codon, initiating translation by coding for methionine, and there are three stop codons (UAG, UAA, UGA) that terminate translation.

During translation, the small ribosomal subunit binds to the mRNA and an initiator tRNA, which adheres to the start codon. The large ribosomal subunit then joins to complete the translation initiation complex. The tRNA corresponding to the next codon enters the ribosome, carrying an amino acid that binds covalently to the methionine from the initiator tRNA. The first tRNA detaches and leaves the ribosome, which shifts to accommodate the next tRNA. This process continues along the mRNA, with tRNAs entering and exiting the ribosome according to the codons on the mRNA, and the polypeptide chain grows. When a stop codon is reached, the completed polypeptide is released, likely entering a cell organelle for folding and further modification \cite{alberts2002molecular}.

\begin{figure}[htbp!]
\begin{center}
    \begin{tikzpicture}[
        adenin/.style = {decoration = {markings,
        mark = at position #1 with { \adenin{(0, 0)}{270} }
  }}, thymin/.style = {decoration = {markings,
        mark = at position #1 with { \thymin{(0, 0)}{270} }
  }}, guanin/.style = {decoration = {markings,
        mark = at position #1 with { \guanin{(0, 0)}{270} }
  }}, cytosin/.style = {decoration = {markings,
        mark = at position #1 with { \cytosin{(0, 0)}{270} }
  }}]
  \newcommand*{\adenin}[2]{\begin{scope}[shift = {#1}, rotate = #2, fill = red]%
        \fill(0, -.1) -- (.35, -.1) -- (.45, 0) -- (.35, .1) -- (0, .1) -- cycle;%
  \end{scope}}%
  \newcommand*{\thymin}[2]{\begin{scope}[shift = {#1}, rotate = #2, fill = cyan]%
        \fill(0, -.1) -- (.35, -.1) -- (.25, 0) -- (.35, .1) -- (0, .1) -- cycle;
  \end{scope}}%
  \newcommand*{\guanin}[2]{\begin{scope}[shift = {#1}, rotate = #2, fill = green]%
        \fill(0, -.1) -- (.35, -.1) arc(-90:90:.1) -- (.35, .1) -- (0, .1) -- cycle;
  \end{scope}}%
  \newcommand*{\cytosin}[2]{\begin{scope}[shift = {#1}, rotate = #2, fill = yellow]%
        \fill(0, -.1) -- (.35, -.1) arc(270:90:.1) -- (.35, .1) -- (0, .1) -- cycle;%
  \end{scope}}%

    \draw[double distance = 1pt, line cap = rect,
        thymin/.list = {132pt,182pt,190pt,227pt},
        adenin/.list = {140pt},
        guanin/.list = {91pt,124pt,153pt,161pt,198pt},
        cytosin/.list = {99pt,107pt,169pt,211pt,219pt},
      preaction = {decorate}, shorten <= -3pt]
        (0, 0.5) .. controls (1.5, 1)  and (2, 1) .. (2.4, 1)
        .. controls (4, 1) and (6, 1) .. (8, 0.4);

    \draw[double distance = 1pt, line cap = rect,
        thymin/.list = {9pt,17pt,38pt},
        adenin/.list = {30pt,46pt,61pt},
        guanin/.list = {1pt,77pt},
        cytosin/.list = {69pt},
      preaction = {decorate}, shorten <= -3pt]
        (0, 0.5) .. controls (1.5, 1)  and (2, 1) .. (2.4, 1)
        .. controls (4, 1) and (6, 1) .. (8, 0.4);

\draw    (-0.2,-0.35) -- (8.08,-0.35) ;

\draw  [color={rgb, 255:red, 126; green, 211; blue, 33 }  ,draw opacity=1 ][fill={rgb, 255:red, 184; green, 233; blue, 134 }  ,fill opacity=1 ] (0,-0.2) -- (0.8,-0.2) -- (0.8,-0.5) -- (0,-0.5) -- cycle ;
\draw    (0,0.05) -- (0,-0.15) ;
\draw    (0.8,0.05) -- (0.8,-0.15) ;
\draw    (0,-0.15) -- (0.8,-0.15) ;
\draw  [color={rgb, 255:red, 20; green, 188; blue, 152 }  ,draw opacity=1 ][fill={rgb, 255:red, 80; green, 227; blue, 194 }  ,fill opacity=1 ] (0.97,-0.2) -- (1.77,-0.2) -- (1.77,-0.5) -- (0.97,-0.5) -- cycle ;
\draw    (0.97,0.05) -- (0.97,-0.15) ;
\draw    (1.77,0.05) -- (1.77,-0.15) ;
\draw    (0.97,-0.15) -- (1.77,-0.15) ;
\draw  [color={rgb, 255:red, 245; green, 166; blue, 35 }  ,draw opacity=1 ][fill={rgb, 255:red, 255; green, 198; blue, 118 }  ,fill opacity=1 ] (1.97,-0.2) -- (2.77,-0.2) -- (2.77,-0.5) -- (1.97,-0.5) -- cycle ;
\draw    (1.97,0.05) -- (1.97,-0.15) ;
\draw    (2.77,0.05) -- (2.77,-0.15) ;
\draw    (1.97,-0.15) -- (2.77,-0.15) ;
\draw  [color={rgb, 255:red, 208; green, 2; blue, 27 }  ,draw opacity=1 ][fill={rgb, 255:red, 254; green, 107; blue, 125 }  ,fill opacity=1 ] (3,-0.2) -- (3.8,-0.2) -- (3.8,-0.5) -- (3,-0.5) -- cycle ;
\draw    (3,0.05) -- (3,-0.15) ;
\draw    (3.8,0.05) -- (3.8,-0.15) ;
\draw    (3,-0.15) -- (3.8,-0.15) ;
\draw  [color={rgb, 255:red, 144; green, 19; blue, 254 }  ,draw opacity=1 ][fill={rgb, 255:red, 187; green, 110; blue, 255 }  ,fill opacity=1 ] (4.14,-0.2) -- (4.94,-0.2) -- (4.94,-0.5) -- (4.14,-0.5) -- cycle ;
\draw    (4.14,0.05) -- (4.14,-0.15) ;
\draw    (4.94,0.05) -- (4.94,-0.15) ;
\draw    (4.14,-0.15) -- (4.94,-0.15) ;
 \draw  [color={rgb, 255:red, 189; green, 16; blue, 224 }  ,draw opacity=1 ][fill={rgb, 255:red, 231; green, 132; blue, 250 }  ,fill opacity=1 ] (5.12,-0.2) -- (5.92,-0.2) -- (5.92,-0.5) -- (5.12,-0.5) -- cycle ;
 \draw    (5.12,0.05) -- (5.12,-0.15) ;
 \draw    (5.92,0.05) -- (5.92,-0.15) ;
 \draw    (5.12,-0.15) -- (5.92,-0.15) ; 
 \draw  [color={rgb, 255:red, 139; green, 87; blue, 42 }  ,draw opacity=1 ][fill={rgb, 255:red, 175; green, 131; blue, 93 }  ,fill opacity=1 ] (6.11,-0.2) -- (6.91,-0.2) -- (6.91,-0.5) -- (6.11,-0.5) -- cycle ;
\draw    (6.11,0.05) -- (6.11,-0.15) ;
\draw    (6.91,0.05) -- (6.91,-0.15) ;
\draw    (6.11,-0.15) -- (6.91,-0.15) ;
\draw  [color={rgb, 255:red, 248; green, 231; blue, 28 }  ,draw opacity=1 ][fill={rgb, 255:red, 254; green, 245; blue, 145 }  ,fill opacity=1 ] (7.08,-0.2) -- (7.88,-0.2) -- (7.88,-0.5) -- (7.08,-0.5) -- cycle ;
\draw    (7.08,0.05) -- (7.08,-0.15) ;
\draw    (7.88,0.05) -- (7.88,-0.15) ;
\draw    (7.08,-0.15) -- (7.88,-0.15) ;

\draw (-1,1) node [anchor=north west][inner sep=0.75pt]   [align=left] {mRNA};

\draw (-1.3,0) node [anchor=north west][inner sep=0.75pt]   [align=left] {protein};
        
\end{tikzpicture}

\caption{Schematic representation of the  RNA translation.}
    \label{fig:translation}    
\end{center}

\end{figure}
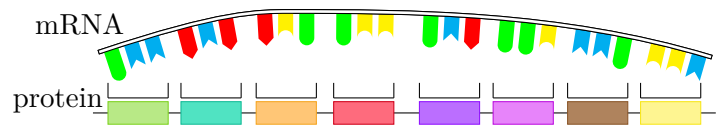

Through this two-step process, DNA is transcribed into mRNA, and then this mRNA is translated into a protein. Since every gene encodes a specific protein and proteins constitute much of an organism's structure and function, from muscle and organ tissue to receptors and enzymes, this is how DNA carries the code for a living organism.

\section{The Andrecut-Kauffman model}
\label{sec:model}

In molecular biology, the flow of genetic information follows this transcription and translation pathway, where DNA is first transcribed into messenger RNA (mRNA), which is subsequently translated into protein. Each gene within DNA encodes a specific protein, essential for an organism's structure and function, forming cells, tissues, and organs, and playing critical roles in processes like muscle structure and enzymatic activity. Thus, DNA serves as the blueprint for life by directing the synthesis of the proteins required for a functioning organism.

\subsection{Motivation behind the model.}
\label{sec:motivation}
Andrecut and Kauffman \cite{andrecut_main} introduced a simplified model of gene expression that merges the processes of transcription and translation described in Section~\ref{sec:mechanisms} into a single reaction.  This approach simplifies the complex biological processes by treating them as one integrated step, making it easier to analyze the model's dynamics.

The main motivation for the introduction of the discrete Andrecut-Kauffman model is the set of chemical reactions for the two-gene system, which covers the complex steps of gene expression. 
In this model, the key reactions include the following.
\begin{enumerate}
    \item [1.] The expression reaction 
    \[
    \textit{RNAp} + P \overset{k}{\longrightarrow} \textit{RNAp} + P + M
    \]
    describes the processes of transcription and translation in a condensed form. During transcription, RNA polymerase ($\textit{RNAp}$) binds to the gene promoter ($P$) and then goes through intermediate reactions, leading to the formation of an mRNA molecule. During translation, the mRNA binds to the ribosome, resulting in the formation of a protein ($M$).
    \item [2.] Multimerization
    \[
        nM \longleftrightarrow M_n
    \]
    is the process by which individual protein particles -- monomers ($n=1$) combine to form a multimeric structure ($M_n$). This step is crucial for many proteins, which achieve full functionality only in their multimeric forms, such as dimers ($n=2$) or trimers ($n=3$), depending on cellular conditions.
    \item [3.] Promotor blocking
    \[
    P + M_n \longleftrightarrow P(M_n)
    \]
    captures how multimeric complexes ($M_n$) bind to promoter sites ($P$), thus
blocking further transcription. Such promoter blocking mechanisms are essential for
regulatory feedback, where the complex ($P(M_n)$) can inhibit continued gene expression.
    \item [4.] The degradation reaction \[
       M  \overset{k^\delta}{\longrightarrow} \text{\O}
       \]
describes the degradation of protein monomers ($M$), a vital aspect of gene regulation that affects the level of gene expression by controlling the lifespan of protein molecules.
\end{enumerate}

\subsection{Equations for the two-gene model.}
\label{sec:equationsOld}
In this framework, the dynamic relationships between the expression levels of two genes, labeled \( x \) and \( y \), are explored through their interactions and feedback mechanisms. The model defines these relationships with the following equations:
\begin{equation}
\label{xy_old}
\left\{\;
\begin{aligned}
x_{t + 1} &= \frac{\alpha}{1+(1-\varepsilon)x^n_t+\varepsilon y^n_t} +\beta x_t,\\
y_{t+1} &= \frac{\alpha}{1+\varepsilon x^n_t+(1-\varepsilon) y^n_t}+\beta y_t.
\end{aligned}
\right.
\end{equation}
Here, \( x_t \) and \( y_t \) represent the concentration levels of transcription factor proteins for the genes \( x \) and \( y \) at time \( t \), respectively. The parameters represent properties of the chemical reactions discussed in Section~\ref{sec:motivation}.
The parameter \( \alpha=\Delta t \cdot k\) reflects the rate of combined transcription and translation over a fixed time interval. The parameter \( \beta = 1-\Delta t\cdot k^{\delta} \) depends on the rate of protein degradation for a fixed time interval. Finally, \( \varepsilon \) is a parameter that describes the coupling between the genes under the assumption that it is symmetric \cite{Andrecut_2006}. The exponent \( n \) denotes the number of monomers of a given protein that are subject to the multimerization reaction. These parameters were considered in \cite{andrecut_main} in the following ranges: $n \in \{1,2,3,4\}$, $\alpha\in[0,100]$, $\beta\in [0,1)$, and $\varepsilon\in [0,1]$.

\subsection{Equations for the generalized two-gene model.}
\label{sec:equations}
The model proposed in \cite{andrecut_main} assumes many simplifications, including the assumption that the genes are identical in terms of the rate of gene expression reaction, and the degradation of protein monomers as well. In \cite{sharma2019,subramani2023}, the Andrecut-Kauffman model was considered with different rates of degradation \( \beta \) for each gene. However, in our research, we will instead deal with different reaction rates $\alpha$ for each gene. In other words, we use $\alpha_1$ and $\alpha_2$ in the two equations instead of having the common rate $\alpha$. We thus consider the following model:
\begin{equation}
\label{xy_model}
\left\{\;
\begin{aligned}
x_{t + 1} &= \frac{\alpha_1}{1+(1-\varepsilon)x^n_t+\varepsilon y^n_t} +\beta x_t, \\
y_{t+1} &= \frac{\alpha_2}{1+\varepsilon x^n_t+(1-\varepsilon) y^n_t}+\beta y_t.
\end{aligned}
\right.
\end{equation}
Moreover, the definition of the parameter $\alpha=\Delta t\cdot k$ in no way limits its value to the range $[0,100]$ considered previously; therefore, in what follows, we consider larger ranges of the parameters $\alpha_1$ and $\alpha_2$ in order to examine the dynamics of the model more broadly.

In the original analysis, Andrecut and Kauffman \cite{andrecut_main} indicate that for $n=1$ and $n=2$ the model exhibits no chaotic behavior and only produces stable points or cycles. We have conducted low resolution simulations to validate these findings using methods described in Section~\ref{sec:LyapComp}. Indeed, the computed Lyapunov exponents were negative in the range of parameters considered. Therefore, we decided to limit our attention to $n=3$ and $n=4$ in our research.

\section{An absorbing set for the model}
\label{sec:absorbing}

Before conducting numerical investigation of model \eqref{xy_model}, it is worth to know that within the regime of parameters of our interest and among all the relevant initial conditions there are no trajectories that become unbounded in forward time. Moreover, we would like to know that all the attractors and other sets that represent long-term dynamics are located within a specific bounded region of the phase space, so that we can conduct informed numerical simulations and not worry that we miss any important part of the dynamics. The following notion is useful for this purpose.

\begin{definition}[Absorbing set, see {\cite[Definition~1]{pilarczyk-graff-2024}}]
A set $P \subset X$ is called an \emph{absorbing set} for a dynamical system generated by $f \colon X \to X$ if for every $x \in X$ there exists $N_0 > 0$ such that $f^n(x) \in P$ for all $n \geq N_0$.
\end{definition}

Note that although every forward trajectory in the dynamical system enters an absorbing set in a finite number of steps, the actual number of steps may depend on the initial conditions. Moreover, every absorbing set contains all the attractors of the system (both local and global). Moreover, it contains all the recurrent dynamics present in the system, such as stable and unstable fixed points and periodic orbits.

Note that the entire phase space satisfies the definition of an absorbing set for a dynamical system. However, we are interested in finding a possibly small bounded absorbing set in order to be able to restrict our numerical simulations to a manageable subset of the phase space and still ensure that we obtain reliable results. The following proposition defines a useful absorbing set for model~\eqref{xy_model}.

\begin{proposition}
\label{prop:abs}
Let $c_1,c_2 > 0$ be arbitrary numbers (in particular, they can be very small). Define $b_i := (\alpha_i + c_i)/(1-\beta)$ for $i \in \{1, 2\}$. Then the set $R_c := [0,b_1] \times [0,b_2]$ is an absorbing set for model \eqref{xy_model} with $\alpha_1, \alpha_2 \geq 0$, $\beta \in [0,1)$, and any $\varepsilon \in [0,1]$ and $n \in \mathbb{N}$.
Moreover, the set $R_c$ is positively invariant.
\end{proposition}

\begin{proof}
Note that only the first quadrant of the phase space (that is, $x,y \geq 0$) is meaningful from the point of view of applications. So we consider $X := \mathbb{R}_+^2$.

Following the ideas from \cite{pilarczyk-graff-2024}, we shall prove that if $x_t \geq b_1$ then $x_{t+1} \leq x_t - c_1$, that is, after each iteration, the $x$ coordinate of the point decreases at least by the constant amount $c_1$, as long as it is beyond $b_1$. This statement implies that after a finite number $N_0$ of iterations, we must reach the condition $x_{t+N_0} \leq b_1$. Moreover, we shall also show that if $x_t \leq b_1$ then also $x_{t+1} \leq b_1$.

For the first statement,
first notice that $x_t \geq b_1 = (\alpha_1 + c_1) / (1 - \beta)$ is equivalent to $\alpha_1 \leq (1 - \beta) x_t - c_1$,
and then calculate as follows:

\begin{equation}
  \begin{split}
x_{t+1} &= \alpha_1 / (1 + \text{a non-negative term}) + \beta x_t
\leq \alpha_1 + \beta x_t \\
&\leq (1 - \beta) x_t - c_1 + \beta x_t = x_t - c_1.
  \end{split}
\end{equation}

Let us now consider the second statement and take $x_t \leq b_1 = (\alpha_1 + c_1) / (1 - \beta)$. Then we calculate:

 \begin{equation}
  \begin{split}
x_{t+1} &\leq \alpha_1 + \beta x_t \leq \alpha_1 + \beta (\alpha_1 + c_1) / (1 - \beta) \\
&= \alpha_1 (1 - \beta) / (1 - \beta) + \beta (\alpha_1 + c_1) / (1 - \beta) \\
&= (\alpha_1 - \beta \alpha_1 + \beta \alpha_1 + \beta c_1) / (1 - \beta) \\
&= (\alpha_1 + \beta c_1) / (1 - \beta)
\leq (\alpha_1 + c_1) / (1 - \beta) = b_1.
  \end{split}
 \end{equation}

By symmetry in the equations, the same holds true for $y_t$ and $b_2$.

As a consequence, every trajectory enters $R_c = X \cap \{x \leq b_1\} \cap \{y \leq b_2\}$ in a finite number of steps. Moreover, every trajectory that starts in $R_c$ remains in $R_c$. This completes the proof.
\end{proof}

In our case, it follows from Proposition~\ref{prop:abs} that if $\alpha_i \in [0,100]$ and $\beta \leq 0.5$ then all the recurrent dynamics of the model we consider is contained in $[0,200]^2$.

\section{Periodic and chaotic behavior of the model}
\label{sec:chaoticperiodic}

Both models \eqref{xy_old} and \eqref{xy_model} admit periodic solutions and chaotic dynamics. However, allowing $\alpha_1 \neq \alpha_2$ may switch the model to quantitatively different dynamics by changing $\alpha_2$ alone while leaving $\alpha_1$ unchanged, as we illustrate below.

\begin{figure}[htbp!]
\begin{center}
     \subfigure[A trajectory for $\alpha_1=\alpha_2=20$.]{\includegraphics[width=0.49\textwidth]{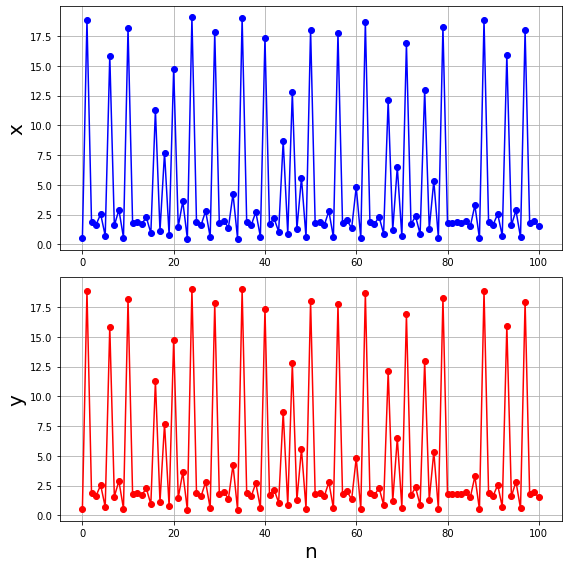}\label{fig:tra_e}} 
    \subfigure[A trajectory for $\alpha_1=20$ and $\alpha_2=60$.]{\includegraphics[width=0.49\textwidth]{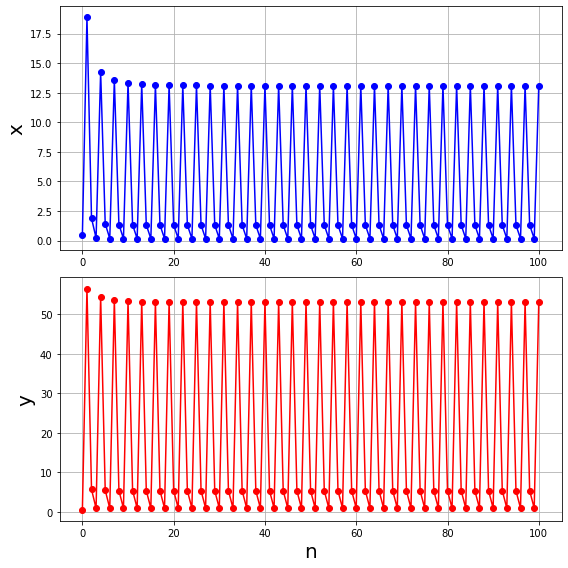}\label{fig:tra_n}} 
    \caption{The two coordinates ($x$ and $y$) of selected trajectories observed for $\varepsilon = 0.8$, $\beta = 0.1$, $n = 4$, and different values of $\alpha_1$ and $\alpha_2$.}
    \label{fig:tra}   
\end{center}
\end{figure}

The plots in Figure \ref{fig:tra} show evolution of variables \(x\) and \(y\) in time in model \eqref{xy_model} for two specific trajectories, both starting at $x_0 = y_0 = 0.5$: one for the model with $\alpha_1=\alpha_2=20$, the other one with the same value of the first reaction rate $\alpha_1 = 20$, but with $\alpha_2 = 60$. In this example, the same values of the other parameters are used in both cases: \(\varepsilon = 0.8\), \(\beta = 0.1\), \(n = 4\). For $\alpha_1=\alpha_2=20$ one can immediately notice the chaotic nature of the model. However, the trajectory for $\alpha_2 = 60$ appears to approach a stable periodic orbit.

The opposite situation is also present in the model. Specifically, it is possible to find a case in which for $\alpha_1=\alpha_2$ we observe periodic solutions, while changing the value of $\alpha_2$ yields chaotic behavior. An example of such a case occurs for the parameters $\alpha_1=3$, $\alpha_2=3$, $\beta=1$, $\varepsilon=0.8$, and $n=4$. Periodic trajectories switch to chaotic solutions when we change the value of $\alpha_2$ to $28$.

These plots highlight the role of \(\alpha_1\) and \(\alpha_2\) in controlling the model's stability, and illustrate how the gene expression model can shift between oscillatory and chaotic regimes, depending solely on the choice of these parameters.

\section{Analysis of bifurcation diagrams}
\label{sec:bifdiag}

A bifurcation diagram is a visual representation of the evolution of the global attractor (or attractors in general) of a dynamical system as a selected one-dimensional parameter varies. The horizontal axis of the diagram typically represents the parameter being varied, while the vertical axis corresponds to the one-dimensional phase space of the system, or a projection onto a selected variable if the actual dimension of the phase space is higher than one. Several features of the global attractor can be easily spotted on a bifurcation diagram, such as the size and complexity of the attractor, as well as bifurcations it undergoes when the parameter is varied.

The purpose of the experiment described below is to investigate the dynamics by means of bifurcation diagrams with respect to one of the reaction rates $\alpha_1$ while keeping the other rate $\alpha_2$ fixed in model \eqref{xy_model} and compare it to the dynamics of model \eqref{xy_old} in which both reaction rates are the same. For that purpose, we fix $\varepsilon = 0.7$, $\beta = 0.2$, $n=4$, $\alpha_2 = 200$, and create bifurcation diagrams for $\alpha_1 \in [0,500]$ for model \eqref{xy_model}. We do the same for model \eqref{xy_old} with $\alpha \in [0,500]$.

\begin{figure}[htbp!]
    \centering
    \subfigure[Bifurcation Diagram for $x$ with $\alpha = \alpha_1=\alpha_2$.]{\includegraphics[width=0.49\textwidth]{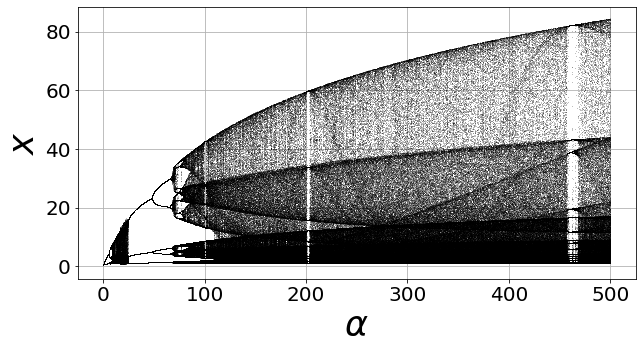}\label{fig:bif_x}} 
    \subfigure[Bifurcation Diagram for $x$ with $\alpha_2=200$.]{\includegraphics[width=0.49\textwidth]{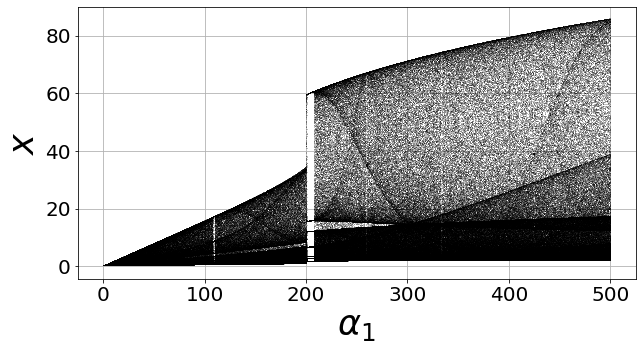}\label{fig:bif_2x}} 
    \subfigure[Bifurcation Diagram for $y$ with $\alpha = \alpha_1=\alpha_2$.]{\includegraphics[width=0.49\textwidth]{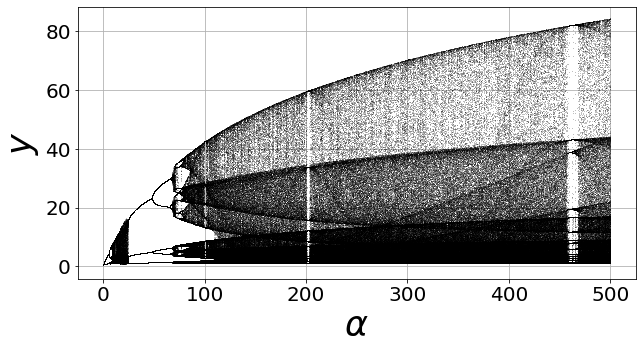}\label{fig:bif_y}}
    \subfigure[Bifurcation Diagram for $y$ with $\alpha_2=200$.]{\includegraphics[width=0.49\textwidth]{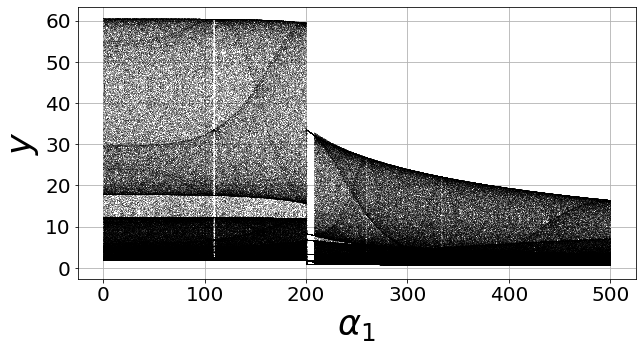}\label{fig:bif_2y}}
    \caption{Bifurcation diagrams of variables $x$ and $y$ for model \eqref{xy_model} with fixed parameters $\varepsilon=0.7$, $\beta=0.2$, $n=4$ for $\alpha_1=\alpha_2$ (a), (c) and $\alpha_2=200$ (b), (d).}
    \label{fig:bif}
\end{figure}

The results of this experiment are shown in Figure~\ref{fig:bif}.
In the vertical axes of the bifurcation diagrams, we show projections of the observed dynamics to variables $x$ and $y$, respectively.
In the horizontal axes, we indicate the parameters of the models that are varied: the common reaction rate $\alpha$ in model \eqref{xy_old} in Figures \ref{fig:bif_x} and \ref{fig:bif_y}, and the reaction rate $\alpha_1$ of model \eqref{xy_model} in Figures \ref{fig:bif_2x} and~\ref{fig:bif_2y}.

It is interesting to see that there is a sudden transition in model \eqref{xy_model} that occurs in the bifurcation diagrams \ref{fig:bif_2x} and \ref{fig:bif_2y} as the parameter $\alpha_1$ crosses the value $\alpha_1 = \alpha_2 = 200$.
Indeed, this is kind of a boundary between two different behaviors of the model. Moreover, we notice considerable difference between the shape of diagrams \ref{fig:bif_2x} and~\ref{fig:bif_2y}. The first one resembles a series of period-doubling bifurcations, while the second one looks like showing period-halving bifurcations.

On the other hand, the corresponding bifurcation diagrams for model \eqref{xy_old} shown in Figures \ref{fig:bif_x} and~\ref{fig:bif_y} for $\alpha$ varying together in the same wide range $[0,500]$ yield relatively typical bifurcation diagrams with periodic windows, period-doubling bifurcations, and regions of allegedly chaotic behavior. Moreover, the diagrams for $x$ and $y$ seem to be practically identical.

\begin{figure}[htbp!]
\centering
\includegraphics[width=0.4\textwidth]{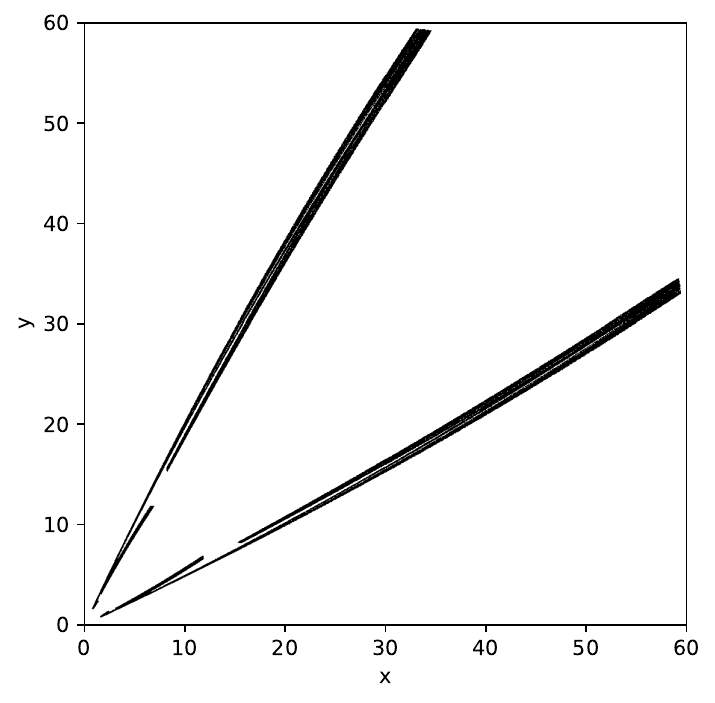}
\caption{Two attractors observed in model \eqref{xy_model} with the parameters $\varepsilon=0.7$, $\beta=0.2$, $n=4$ and $\alpha_1=\alpha_2=200$.}
\label{fig:attr200}
\end{figure}

In an attempt to seek the reasons for the $x$-$y$ symmetry in the bifurcation diagrams shown in Figures \ref{fig:bif_x} and~\ref{fig:bif_y}, as well as the odd bifurcation observed in Figures \ref{fig:bif_2x} and~\ref{fig:bif_2y} at $\alpha_1 = 200$, we used numerical simulations to compute the actual attractors in the phase space for several values of $\alpha_1$. It turns out that for $\alpha_1 = \alpha_2 = 200$ one can see two attractors symmetric to each other, as shown in Figure~\ref{fig:attr200}. When we increase or decrease $\alpha_1$ by a certain amount, while keeping the other parameters unchanged, including $\alpha_2 = 200$, one of the attractors disappears and the model exhibits a single attractor. The symmetry of equations \eqref{xy_old} and the existence of the two attractors explain the fact that we observed identical bifurcation diagrams in Figures \ref{fig:bif_x} and~\ref{fig:bif_y}. The disappearance of one of the attractors, on the other hand, explains the sudden change in the shape of the observed projections of the attractors in Figures \ref{fig:bif_2x} and~\ref{fig:bif_2y} when $\alpha_1$ crosses the value of $\alpha_2 = 200$.

\section{Numerical procedure for the computation of the maximum Lyapunov exponent}
\label{sec:LyapComp}

Chaos is characterized by aperiodic long-term behavior of a bounded deterministic system that exhibits sensitive dependence on initial conditions. In such a system, small differences in initial conditions can lead to vastly different outcomes, making prediction difficult. A key tool for quantifying the separation rate of such trajectories is the Lyapunov exponent. In general, the existence of a positive Lyapunov exponent indicates instability and chaotic behavior, while having all the Lyapunov exponents negative suggests convergence of nearby trajectories and thus stability. In an $n$-dimensional system, there are $n$ Lyapunov exponents corresponding to the system's degrees of freedom. See, e.g., \cite{sprott2003chaos} for a comprehensive introduction to this topic.

Let us explain the background and the numerical approach that we use to compute the maximum Lyapunov exponent for our model.

In what follows, we are interested in the maximum Lyapunov exponent. The maximum Lyapunov exponent is simply the larger of the two Lyapunov exponents in our two-dimensional model.

\begin{definition} [see \cite{Kathleen-1996}]
Let $f$ be a smooth map on $\mathbb{R}^2$, and let $J_m=Df^m(v_0)$ denote the first derivative matrix of the $m$-th iterate of $f$ at the initial point $v_0$. For $i\in\{1,2\}$, let $r^m_i$ be the length of the $i$-th longest orthogonal axis of the ellipsoid $J_m N$ for an orbit with initial point $v_0$. Then $r^m_i$ measures the contraction or expansion near the orbit of $v_0$ during the first $m$ iterations. The $i$-th \emph{Lyapunov number} of $v_0$ is defined by
\[
L_i=\lim_{m\rightarrow\infty}\left(r^m_i\right)^{1/m},
\]
if this limit exists. The corresponding $i$-th \emph{Lyapunov exponent} of $v_0$ is $\lambda_i=\ln L_i$.
\end{definition}

In order to compute the maximum Lyapunov exponent for model \eqref{xy_model}, the Jacobian matrix of the right-hand side is determined as follows:
\begin{equation}
\renewcommand\arraystretch{2.5}
 J=  \begin{bmatrix}
-\dfrac{(1-\varepsilon)\alpha_1nx^{n-1}}{\left(1+(1-\varepsilon)x^n+\varepsilon y^n\right)^2} +\beta & -\dfrac{\varepsilon\alpha_1ny^{n-1}}{\left(1+(1-\varepsilon)x^n+\varepsilon y^n\right)^2}\\
-\dfrac{\varepsilon\alpha_2nx^{n-1}}{\left(1+\varepsilon x^n+(1-\varepsilon)y^n\right)^2} & -\dfrac{(1-\varepsilon)\alpha_2ny^{n-1}}{\left(1+\varepsilon x^n+(1-\varepsilon)y^n\right)^2} +\beta
\end{bmatrix} .
\end{equation}
Then an initial small disturbance is iterated along the trajectory, following the equations, using the Jacobian matrix evaluated at consecutive points of the trajectory. After each iteration, the size of the disturbance is measured. It shows how quickly nearby trajectories diverge or converge. A positive value of the maximum Lyapunov exponent suggests chaotic behavior of the model, as it indicates an average exponential divergence of nearby trajectories.

Let us now describe the numerical procedure that we employed to estimate the maximum Lyapunov exponent for a considerable number of points in the parameter space.
The algorithm used during these calculations relies on the fact that for a given point $x$ orbiting inside an attractor, almost every vector in the neighborhood of $x$ will converge towards the direction of the fastest expansion.

In order to calculate the Lyapunov exponent numerically for initial conditions $x_0$ (preferably in a neighborhood of an attractor), a second point $x_0'$ is chosen at a small distance $d_0 = 10^{-8}$ from $x_0$. After one iteration, we obtain $d_1 = \|f(x_0)-f(x_0')\|$. From the definition of the Lyapunov exponents, we expect $d_1 \approx d_0 e^\lambda$, allowing us to estimate $\lambda\approx\ln\frac{d_1}{d_0}$. This process is repeated iteratively with $x_{m+1} = f(x_m)$ and $x_{m+1}' = x_{m+1} + \frac{d_0}{d_{m+1}}(f(x_m') - x_{m+1})$. Normalization is applied at each step to keep both trajectories close enough so that $f$ can be reasonably approximated by its Jacobian, yet sufficiently separated to minimize numerical errors that would make their images merge into one number. The final estimate is the average of the $\lambda$ values obtained in all iterations.

Note that we treat the model as if it were effectively one-dimensional, because the vector $x_m'-x_m$ naturally converges towards the direction of fastest expansion, which aligns with the eigenvector associated with the largest eigenvalue of the Jacobian matrix $J$. This approach does not yield the full Lyapunov spectrum. However, it is inexpensive computationally and the estimated maximum exponent suffices to identify chaotic dynamics in the model.

\begin{figure}
    \centering
    \includegraphics[width=0.8\linewidth]{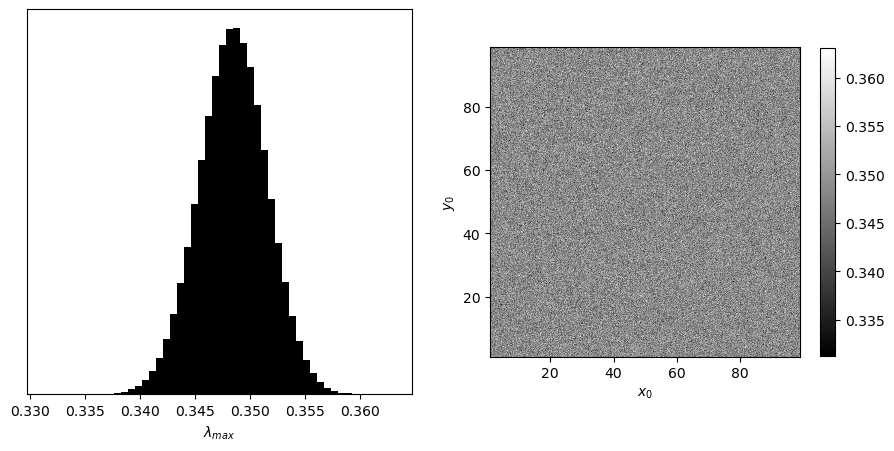}
    \caption{Distribution of the computed Lyapunov exponent for $1024 \times 1024$ starting points taken from $(1, 99) \times (1, 99)$. Parameters used in this simulation are $\alpha_1 = 50$, $\alpha_2 = 85$, $\beta = 0.2$, $\varepsilon = 0.7$ and $n = 3$.}
    \label{fig:lyapunov_histogram}
\end{figure}

If a global attractor exists, the method should converge to a single Lyapunov exponent value, independent of the chosen initial condition. This behavior has been confirmed in the current setup, as shown in Figure~\ref{fig:lyapunov_histogram}. The left panel displays the distribution of Lyapunov exponent values calculated for a large number of initial conditions, showing a normal distribution centered around approximately $0.35$, which supports convergence towards a single value. The right panel shows the spatial distribution of these values in the phase space, where the uniformity and homogeneity of the values provide further evidence of their independence with respect to the choice of $x_0$ and $y_0$. Together, these visualizations provide a baseline for understanding the computational methods discussed in the following sections.

\section{A search for positive Lyapunov exponents in both models}
\label{sec:posLyap}

The purpose of our first numerical experiment that involves the computation of the maximum Lyapunov exponent is to check whether taking different reaction rates $\alpha_1$ and $\alpha_2$ for each gene proposed in model \eqref{xy_model}, as opposed to using the same rate $\alpha = \alpha_1 = \alpha_2$ in model \eqref{xy_old}, increases the number of parameters $(\beta,\varepsilon)$ for which a positive maximum Lyapunov exponent can be encountered.

For that purpose, we took a grid of $100 \times 100$
different values of the parameters \(\beta \in [0,1) \) 
and \(\varepsilon \in [0,1]\). Specifically, we examined $100$ evenly spaced values of $\beta$, ranging from $0$ to $0.99$, including both endpoints, and $100$ values of $\varepsilon$ ranging from $0$ to $1$, inclusive.
For each of these combinations, we computed the maximum Lyapunov exponent for a grid of
$1000 \times 1000$ values of $\alpha_1, \alpha_2 \in [0,100]$, including both $0$ and $100$,
taking the initial condition
$x_0 = 0.1$, $y_0 = 0.1$ in each case. We conducted these computations for each of the two models \eqref{xy_old} and \eqref{xy_model}, and for $n \in \{3, 4\}$, which resulted in four series of computations in total.

\begin{figure}[htbp!]
    \centering
    \subfigure[Results for $n=3$.]{\includegraphics[width=0.49\textwidth]{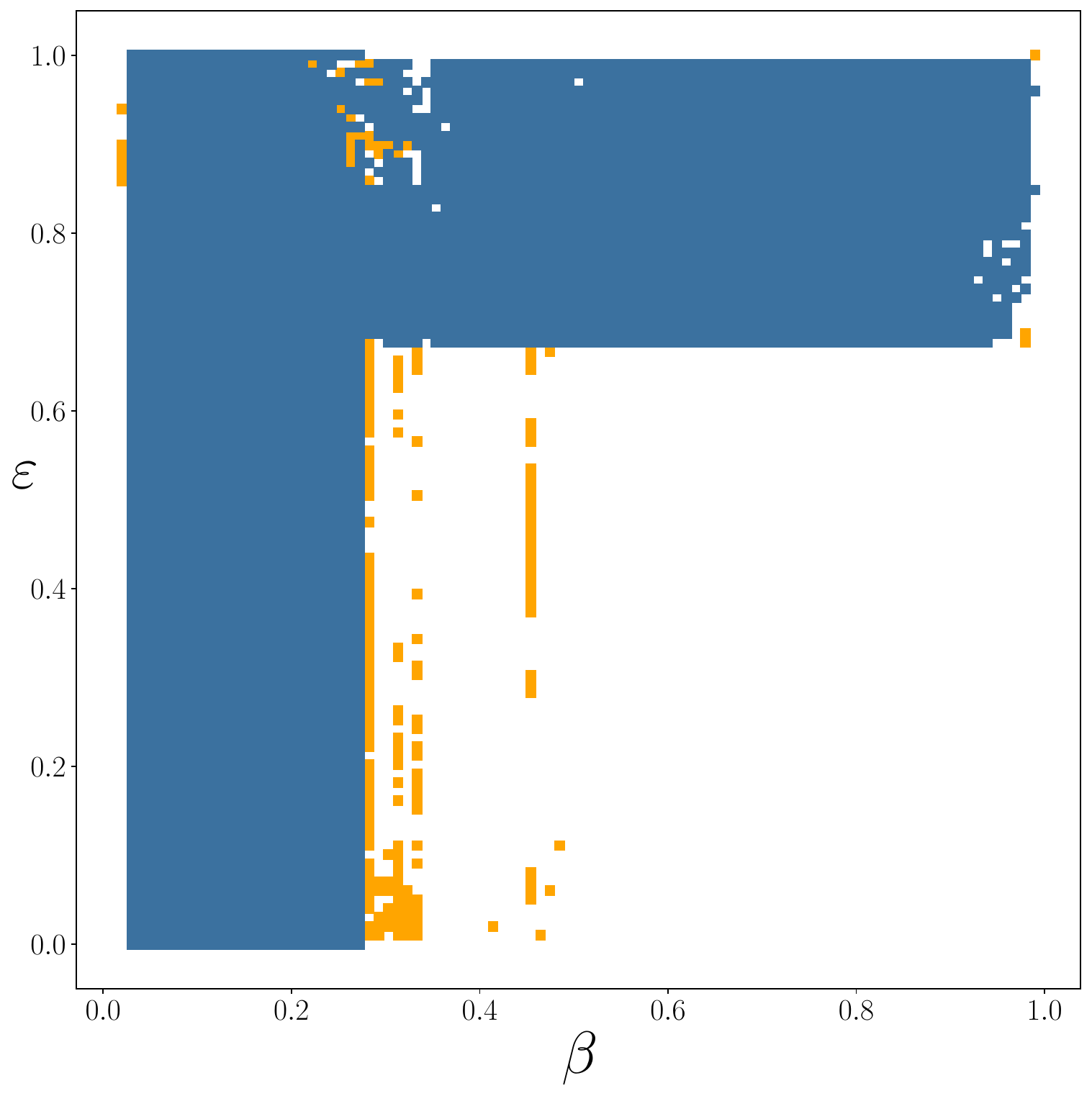}} 
    \subfigure[Results for $n=4$.]{\includegraphics[width=0.49\textwidth]{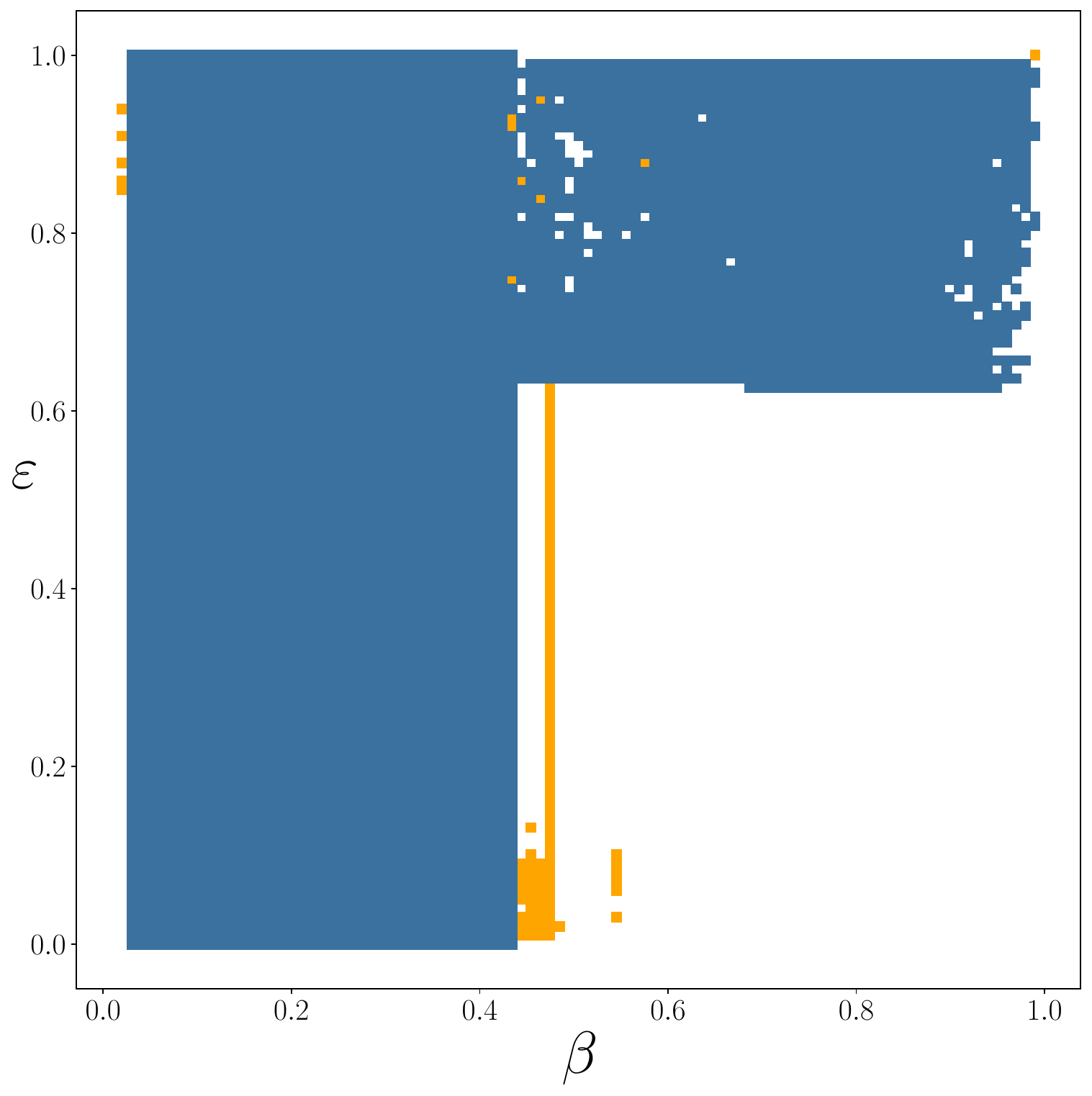}}
    \caption{Pairs of parameters $\beta$ and $\varepsilon$ for models \eqref{xy_old} and \eqref{xy_model} for which there exists a parameter $\alpha\in [0,100]$, or a pair of parameters $(\alpha_1,\alpha_2)\in[0,100]\times[0,100]$, respectively, with the positive maximum Lapunov exponent. Pairs with the positive maximum Lyapunov exponent observed in both models are marked in blue, while those with this observation made only for model \eqref{xy_model} are marked in orange.}
    \label{fig:lyap}
\end{figure}

The plots in Figure~\ref{fig:lyap} illustrate those pairs of parameters \(\beta\) and \(\varepsilon\) for which we were able to find reaction rates \((\alpha_1, \alpha_2) \in [0, 100] \times [0, 100]\) in the case of model \eqref{xy_model}, or a reaction rate \(\alpha \in [0, 100]\) in the case of model \eqref{xy_old}, with a positive maximum Lyapunov exponent. Note that the presence of a positive Lyapunov exponent indicates chaotic behavior encountered in the model.

The plots show that a larger number of parameter pairs \((\beta, \varepsilon)\) for model \eqref{xy_model} were found to be capable of experiencing chaotic behavior, compared to model \eqref{xy_old}. This suggests that model \eqref{xy_model} is more prone to experience chaotic dynamics.

\section{The maximum Lyapunov exponent for various parameters $\alpha_1$ and $\alpha_2$}
\label{sec:plotLyap}

\begin{figure}
    \centering
    \includegraphics[width=0.5\linewidth]{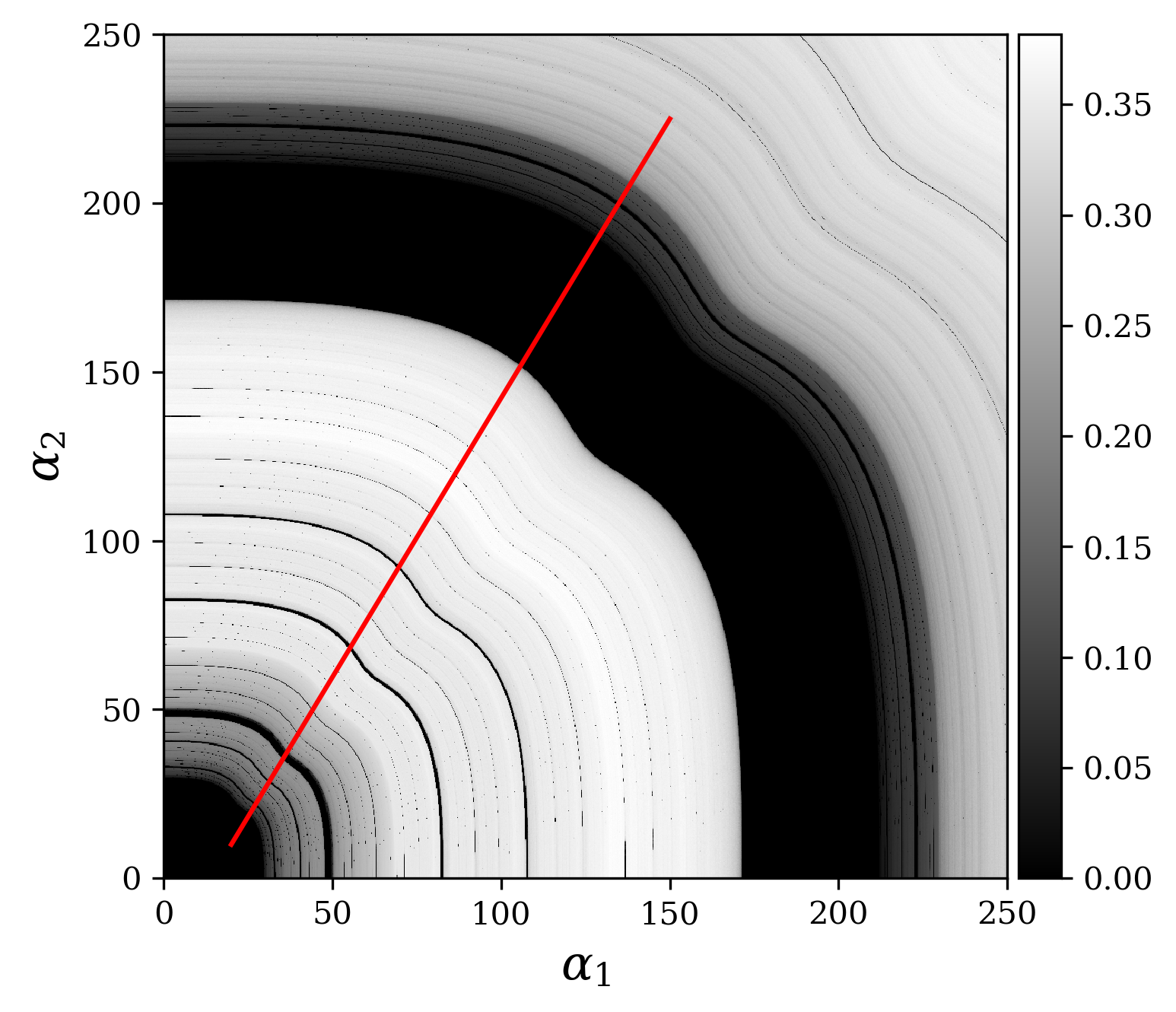}
    \caption{The maximum Lyapunov exponent computed for model \eqref{xy_model} for $\alpha_1, \alpha_2 \in (0, 250)$ with $\beta = 0.2$, $\varepsilon = 0.6$, and $n = 3$.}
    \label{fig:lyapunov_param_slice}
\end{figure}

Plots of Lyapunov exponents as a function of parameters of a model convey detailed information on the changes in the behavior of the model when the parameters vary. In Figure \ref{fig:lyapunov_param_slice}, we show the maximum Lyapunov exponent computed for model \eqref{xy_model} when $\alpha_1$ and $\alpha_2$ vary within the interval $(0, 250)$ while $\beta = 0.2$, $\varepsilon = 0.6$, and $n = 3$ are fixed.

\begin{figure}
    \centering
    \includegraphics[width=0.8\linewidth]{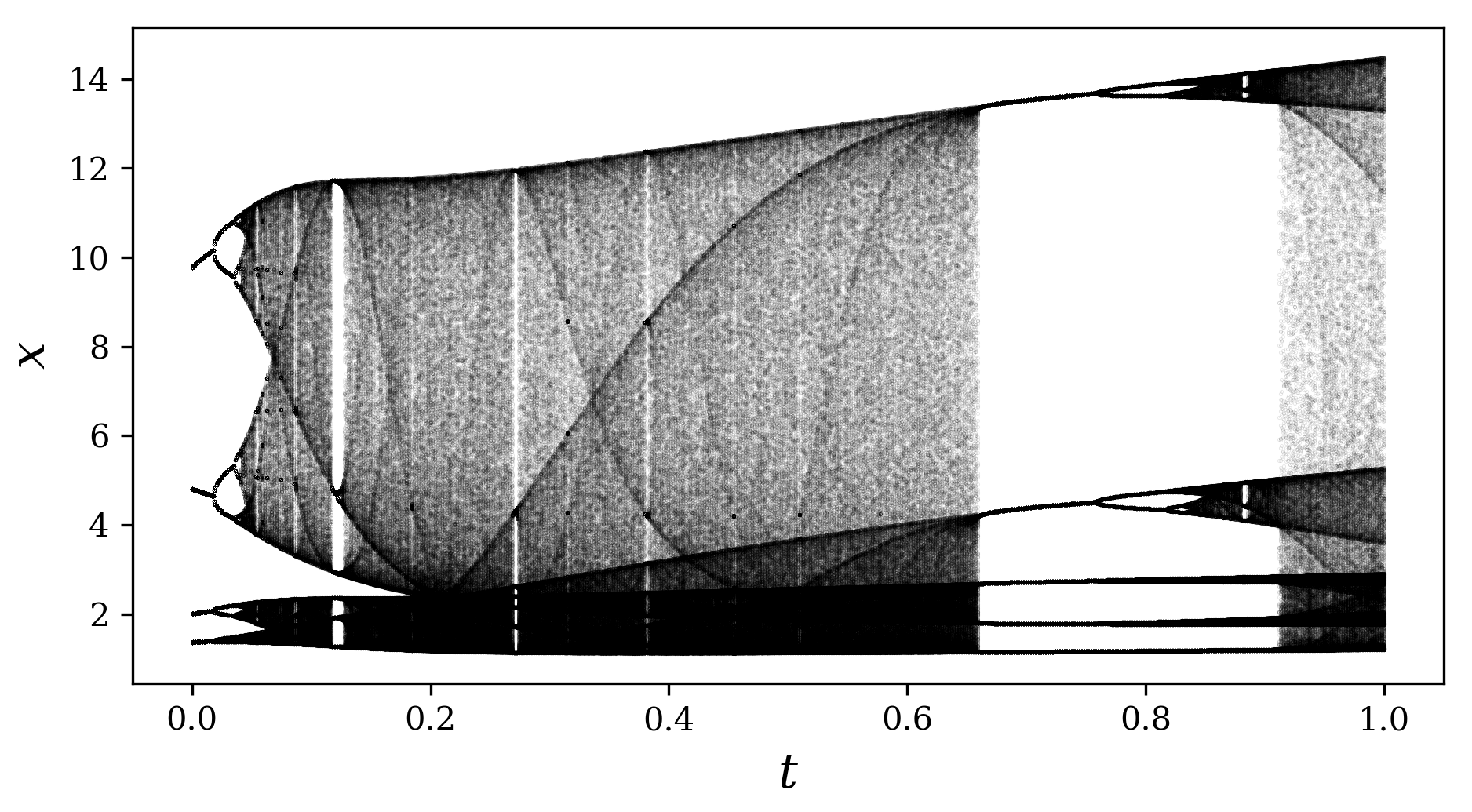}
    \caption{Bifurcation diagram of the $x$ variable computed along the segment in the parameter space plotted in red in the previous figure (Figure~\ref{fig:lyapunov_param_slice}), parametrized as $(\alpha_1, \alpha_2)=(20+130t, 10+215t)$ for $t \in [0,1]$.}
    \label{fig:bifurcation_slice}
\end{figure}

Due to the color scale chosen in Figure \ref{fig:lyapunov_param_slice}, stability regions are shown in black (low values of the maximum Lyapunov exponent), and chaotic dynamics is indicated by lighter shades of gray, with brightness corresponding to the magnitude of chaos, measured by the value of the maximum Lyapunov exponent. As one can see in the figure, moving across the parameter space along a straight line might involve multiple transitions between chaos and stability. Indeed, the bifurcation diagram in Figure~\ref{fig:bifurcation_slice} was computed for the parameters taken along the line segment plotted in red in Figure~\ref{fig:lyapunov_param_slice}. It shows how chaos arises in the model from a series of period-doubling bifurcations at small values of $\alpha_1$ and $\alpha_2$. One can also see the next major stable region that appears when the chaotic attractor collapses into a period-five orbit.

\section{Asymmetry in the computation of the maximum Lyapunov exponent}
\label{sec:asymmetry}

The symmetry in the equations that define this dynamical system implies that, for a global attractor, Lyapunov exponent values in the $(\alpha_1,\alpha_2)$ parameter plane should exhibit symmetry around the $\alpha_1=\alpha_2$ line, as demonstrated in Figure \ref{fig:lyapunov_param_slice}. However, while scanning the parameter space using a fixed set of $16$ initial conditions, we identified regions where that symmetry was unexpectedly broken. This phenomenon can be explained by the presence of multiple attractors in the phase space.

In order to illustrate this phenomenon, we consider a fixed initial condition $x_0$, and let $A_1$ and $A_2$ be two attractors of the model that are present for certain values of $\alpha_1$ and $\alpha_2$. Assume $x_0$ is in the basin of attraction $B_1$ associated with $A_1$. Now consider the model with the parameters $\alpha_1$ and $\alpha_2$ swapped, that is, $\alpha_1' = \alpha_2$ and $\alpha_2' = \alpha_1$. The new model will behave identically to the original one, except that the phase space will be reflected by the $y=x$ line. Consequently, attractors $A_1' = T(A_1)$ and $A_2' = T(A_2)$ will now exist, where $T\colon(x, y)\mapsto(y, x)$. Their basins of attraction will similarly be reflections $B_1' = T(B_1)$, $B_2' = T(B_2)$ of the original ones, and thus it is now possible for $x_0$ to lie within $B_2'$. Although the attractors' locations have changed, their underlying properties, including their Lyapunov spectra, remain the same.

Therefore, if the maximum Lyapunov exponent was $\lambda_1$ in $A_1$ and $\lambda_2$ in $A_2$, with $\lambda_1\neq\lambda_2$ the maximum Lyapunov exponent calculated at the point $x_0$ may differ when the parameters $\alpha_1$ and $\alpha_2$ are swapped, despite the intrinsic symmetry of the equations that define the model.

\begin{figure}
    \centering
    \includegraphics[width=0.7\linewidth]{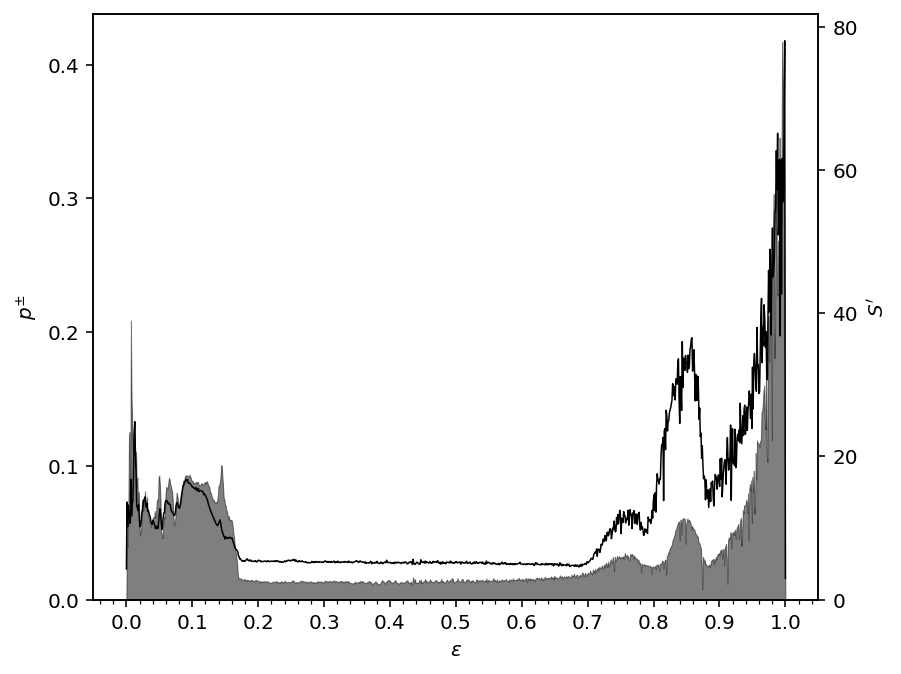}
    \caption{Asymmetry of the model assessed for $\alpha_1, \alpha_2 \in (0, 250)$, $\beta = 0.2$ and $n = 3$ measured by the two quantities introduced in the text: $S'$ shown as the black line and $p^{\pm}$ indicated by the shaded area.}
    \label{fig:symmetry}
\end{figure}

This behavior can be utilized to detect ranges of parameters where certain forms of bistability occur. The following computations focused on the effects of choosing $\varepsilon$. We sampled $1024$ uniformly spaced values of $\varepsilon$ from $[0, 1]$. For each sample, the maximum Lyapunov exponent was calculated for $(\alpha_1, \alpha_2)$ in a $512 \times 512$ grid over the range of $(0, 250)$, resulting in a square array of exponents denoted by $L$. The amount of asymmetry $S'$ was then calculated as
\begin{equation*}
    S' = \sqrt{\sum_{i=1}^{512}\sum_{j=1}^{512} (L_{ij} - L_{ji})^2}
\end{equation*}
and averaged over the 16 initial conditions considered. The computed values of the amount of asymmetry $S'$ are shown in Figure \ref{fig:symmetry} (line graph) together with the proportion of samples for which both positive and negative maximum Lyapunov exponents have been detected (depending on the initial condition chosen), denoted as $p^{\pm}$ (shaded area). Note that the absolute value of $S'$ is less important than its relative magnitude compared to the baseline.

Two intervals with high asymmetry values can be noticed. The results for $\varepsilon\in(0, 0.16)$ appear to reflect slow convergence of the model with a low but nonzero coupling between the variables, and are thus inconclusive. However, the method converges relatively quickly for $\varepsilon\in(0.7, 1)$. Indeed, closer analysis of parameters from the latter subset reveals the presence of bistability which we discuss in the next section.

\section{Three kinds of bistability in the model}
\label{sec:bistability}

For wide ranges of parameters, both models \eqref{xy_old} and \eqref{xy_model} exhibit the phenomenon of bistability. It is defined as the existence of two coexisting attractors, sometimes with different characteristics, such as a periodic orbit or a chaotic attractor. Any trajectory in the model then converges to one attractor or to the other one, which is reflected in the model experiencing a different type of long-term behavior, depending on the initial conditions.

\begin{figure}[htbp!]
\centering
\includegraphics[width=0.49\textwidth]{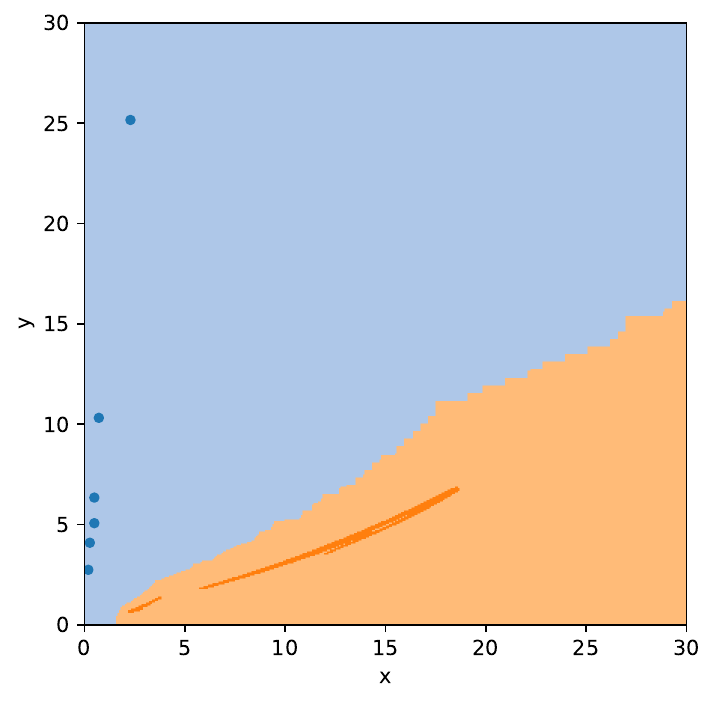}
\caption{An attracting cycle (blue) and a chaotic attractor (orange) observed in model \eqref{xy_model} with the parameters $\alpha_1 = 43.6$, $\alpha_2 = 75.7$, $\varepsilon=0.9$, $\beta=0.2$, and $n=3$ mentioned in Table~\ref{tab:param_values}. Approximations of attraction basins of the attractors are indicated by the corresponding bright colors.}
\label{fig:basins4}
\end{figure}

Examples of the three kinds of bistability that we found are provided in Table~\ref{tab:param_values}.
In particular, the stable period-6 orbit and the chaotic attractor observed numerically for the parameters listed in the third row of Table~\ref{tab:param_values} are shown in Figure~\ref{fig:basins4} together with approximations of their basins of attraction.

\begin{table}[h]
    \caption{Examples of parameters resulting in different modes of bistability of model \eqref{xy_model}.}
    \centering
    \begin{tabular}{c|ccccc}
        type of bistability & $\alpha_1$ & $\alpha_2$ & $\beta$ & $\varepsilon$ & $n$ \\
        \hline\hline
        two attracting cycles & 50.0 & 55.0 & 0.2 & 0.804 & 3 \\
        two chaotic attractors & 55.0 & 55.0 & 0.2 & 0.770 & 3 \\
        an attracting cycle and a chaotic attractor & 43.6 & 75.7 & 0.2 & 0.900 & 3
    \end{tabular}
    \label{tab:param_values}
\end{table}

To the best of our knowledge, this phenomenon was not previously observed for such wide ranges of parameters in the Andrecut-Kauffman model. 
Up to now, the only kind of bistability known for this model was the two stable fixed points present for $\varepsilon = 1$ (see \cite{Andrecut_2006}).

A possible reason for the fact that the bistability of this kind was not discovered in this model by previous studies is that while multiple attractors exist for $\alpha_1 = \alpha_2$, their maximum Lyapunov exponents are identical and therefore cannot be detected by this feature alone. Therefore, admitting different rates $\alpha_1 \neq \alpha_2$ in the model allows for better understanding of possible dynamics of the actual gene expression system.

Finally, it turns out that for $\varepsilon = 0$, the existence of more than two attractors is possible. The variables $x$ and $y$ are independent in that case, and each of them might have its own attractor. In particular, at some values of the parameters, both may have a stable period-4 cycle; let us denote these cycles by $(x_1, x_2, x_3, x_4)$ and $(y_1, y_2, y_3, y_4)$, respectively. Then the Cartesian product of the two 4-element sets is an attractor for the two-dimensional model. It splits into four disjoint attracting cycles:
\begin{align*}
(x_1, y_1)\to(x_2, y_2)\to(x_3, y_3)\to(x_4, y_4)\\
(x_1, y_2)\to(x_2, y_3)\to(x_3, y_4)\to(x_4, y_1)\\
(x_1, y_3)\to(x_2, y_4)\to(x_3, y_1)\to(x_4, y_2)\\
(x_1, y_4)\to(x_2, y_1)\to(x_3, y_2)\to(x_4, y_3)\\
\end{align*}
For example, one can see this behavior for $\alpha_1=13$, $\alpha_2=13$, $\beta=0.2$, $\varepsilon=0$, and $n=3$.

\section{Conclusions and future research directions}
\label{sec:conclusion}

Gene expression is a fundamental biological process that influences nearly every aspect of cell function, including cell differentiation, responses to stimuli, and cell cycle regulation. Research on the dynamics of gene expression provides insight into mechanisms that lead to stable and chaotic patterns of activity, which is crucial in both systems biology and synthetic biology, where the goal is to design controlled genetic systems with specific properties. 

Our results illustrate different types of dynamical behavior encountered across different parameter regimes. These results shed light on the influence of parameters on stability of the model, and thus on the modeled system, and its potential for exhibiting chaotic behavior as well.
In particular, the existence of configurations with the chaos-order type of bistability shown in Section~\ref{sec:bistability} has profound implications for the real gene expression system. A sudden external impulse (e.g., an injection containing one of the transcription factors) or continuous addition of one protein into the system can permanently destabilize the system by pushing it across the boundary between basins of attraction towards the chaotic region. In the same way, ordered state can be reestablished by applying some force towards the stable attractor until the system switches its state. Similar switching between stable states could also occur naturally during processes analogous to cell specialization or cells working in different modes.

The process of multimerization is essential for the functionality of many proteins, as they achieve full functionality only in their multimeric forms, such as dimers ($n = 2$), trimers ($n = 3$) or higher-order multimers, depending on cellular conditions.
Therefore, there is a natural demand for further research on model \eqref{xy_model} to include the analysis of the dynamics also for $n > 4$.
As one can see in Figure~\ref{fig:lyap}, there are more parameters that exhibit chaotic dynamics for $n=4$ than for $n=3$. Therefore, one can expect that the analysis of dynamics becomes more demanding with the increase in~$n$.

Another promising direction of further research might be to search for unstable fixed points and unstable periodic orbits, especially in the case of bistability.
Although biological systems typically settle down in stable states, one can apply mild control (e.g., in terms of some external force or an agent) in order to keep the system in an unstable state at minimal cost. Therefore, determining unstable fixed points and unstable periodic solutions is also important for a better understanding of such systems and handling them.

In general, the analysis of bifurcations and unstable fixed points in these types of models is of particular importance, as it allows the identification of conditions that lead to a qualitative change in the behavior of the system, such as transitions between activated and silenced states of genes.
Thorough analysis of the actual bifurcations that take place in the system when the parameters change might shed light on the actual role of the various parameters for the dynamics exhibited by the system and on the stability of solutions.
As a result of a bifurcation, some stable states might lose their stability.
This means that for the gene expression process, a gene that was previously active can become silenced, potentially leading to changes in cellular behavior or function.
On the other hand, a new stable solution might emerge.
This could correspond to the appearance of a new biological state or pathway that was not previously observed, such as the activation of alternative gene expression mechanisms or regulatory networks.
It might also be the case that chaotic solutions emerge, which means that due to the change of some conditions, such as the increase in temperature that affects the intensity of some processes, the system suddenly switches its state from a well-understood stable equilibrium to hard-to-predict chaotic oscillations.
From the perspective of gene expression, such chaotic behavior could result in irregular production of proteins, leading to unpredictable cellular responses and potentially impacting the organism's overall functionality. 
Both silencing and overexpression as defects in gene regulation can cause profound and unfavorable change in cellular functioning; see, e.g., \cite{li-cao-li-jin-2018,liu-dai-du-2015}.
As a result, many kinds of bifurcations are important from the biological point of view and should not be neglected.

Finally, comprehensive analysis of global dynamics that takes into account all the possible values of parameters in the ranges indicated in Section~\ref{sec:model}, focused on detecting interesting phenomena like bistability or chaotic dynamics, might provide information on how often these phenomena occur and how likely they are to be encountered in the model. For example, one could try applying the approach introduced in \cite{arai-2009} and further developed in \cite{pilarczyk-2023} for this purpose, combined with additional analyses, such as the calculation of the maximum Lyapunov exponent discussed in Section~\ref{sec:LyapComp}. Although this approach is limited to the analysis of dynamics at finite resolution, it can be argued that phenomena occurring below certain scale in a mathematical model of a biological system are not biologically relevant; see, e.g., the discussion in~\cite{LuzzattoPilarczyk2011}.

\section*{Acknowledgments}
The authors M.\ Rosman and A.\ Bartłomiejczyk acknowledge support obtained from Gdańsk University of Technology in the framework of the DEC-27/1/2022/ IDUB/III.4c/Tc grant under the Technetium Talent Management Grants -- ‘Excellence Initiative -- Research University’ program. The authors M.\ Palczewski and P.~Pilarczyk acknowledge support received from the National Science Centre, Poland, within the grant OPUS 2021/41/B/ST1/00405.

\providecommand{\doititle}[1]{#1}

\medskip

\end{document}